\documentstyle[epsfig]{article}
\makeatletter
\let\chapter\hid@chapter
\makeatother
\def\lsim{\lower.5ex\hbox{$\; \buildrel < \over \sim \;$}}
\def\gsim{\lower.5ex\hbox{$\; \buildrel > \over \sim \;$}}
\begin{document}
\pagenumbering{arabic}
\title{First VLF detections of ionospheric disturbances due to Soft Gamma ray Repeater SGR J1550-5418 and Gamma Ray Burst GRB 090424}

\author{Sandip K. Chakrabarti
\footnote{Corresponding Author} \footnote{Also at Indian Centre for Space Physics,
43 Chalantika, Garia Station Road, Kolkata-700 084}
\\
S.N. Bose National Centre for Basic Sciences,\\
JD-Block, Sector-III, Salt Lake, Kolkata-700 098, INDIA\\
Sushanta K. Mondal, Sudipta Sasmal and Debashis Bhowmick\\
Indian Centre for Space Physics, 43 Chalantika, Garia Station Road, Kolkata 700 084\\
Asit K. Choudhury\\
Indian Centre for Space Physics, Malda Branch, Atul Market, Malda 732 101\\
Narendra N. Patra\\
NCRA/TIFR, Ganeshkhind, Pune-411 007, Maharastra, India\\
E-mail: chakraba@boson.bose.res.in
}

\maketitle

\baselineskip 24pt

\begin{abstract}

\noindent {\bf Abstract :} We present the first report of the detection of sudden ionospheric disturbances (SIDs) due to a Soft 
Gamma Ray Repeater (SGR) SGR J1550-5418 and a Gamma Ray Burst (GRB)  GRB 090424. These detections were 
made with receiving stations of Indian Centre for Space Physics which were monitoring Very Low Frequency
signals (VLFs) from the VTX transmitter located at the southern tip of Indian sub-continent. These
positive detections add to the list of a handful of similar detections of other GRBs and SGRs throughout the
world.

\end{abstract}



\noindent {\it Received October 20, 2010, accepted November 3, 2010}

\baselineskip 24pt

\section{Introduction}

It is well known that the earth's ionosphere is a gigantic detector of high energy phenomena
which are taking place in the Universe. The activities such as solar flares or gamma ray bursts
or other such energetic events cause ionospheric disturbances which may be detected 
by studying the  very low frequency (VLF) radio signals which propagate inside the waveguide formed between the 
ionosphere and surface of the earth. While there are many solar flares, particularly, during the 
solar maximum, there are only handful of cases where the gamma ray bursts (GRBs) or soft gamma ray repeaters (SGRs)
have been detected. This is because the GRBs are usually cosmological and the SGRs may also be 
very far out ($\sim 10$kpc) even if in our own galaxy. 

Indian Centre for Space Physics (ICSP) have been monitoring the ionospheric disturbances for about a decade.  
The goals are to study solar flares [1], meteor showers [2], lithosphere-ionosphere 
coupling and precursors to earthquakes [3-6]
and, of course, sudden ionospheric disturbances due to high energy phenomena such as Gamma Ray Bursts 
and Soft Gamma ray Repeaters [7-10]. 
The latter study is very relevant not only to understand the ionospheric chemistry in presence of high energy photons,
it also gives us an idea of how vulnerable the biosphere of the earth is.

In this {\it Rapid Communication}, we report the convincing VLF detection of the 
very energetic soft gamma ray repeater SGR J1550-5418 which erupted several hundred times
on 22nd January, 2009. To our knowledge, ICSP detection was the first such detection [7-8].
Subsequently, ICSP VLF monitors 
also had first positive detections of two bright GRBs [9, 11]. A brief
review of the subject has been presented in Ref. [10].

Ionospheric disturbance due to a GRB named GRB 830801 (GB 830801) was first observed on August 1, 1983
through simultaneous changes in VLF amplitudes observed at Palmer, Antarctica 
receiving station from a transmitter at Rugby, England; Annapolis, 
Maryland and at Lualualei, Hawai [12]. 
Total fluence of the burst was 2$\times 10^{-3}$ erg cm$^{-2}$. 
On August 27th, 1998 at 10:22 UT, the Soft Gamma Ray Repeater 
SGR 1900+14 ionized the exposed part of Earth's night side lower ionosphere. The 
object is considered to be a `magnetar' 12-15 kpc away. Most of the high energy
satellites were saturated [13]. The Tokyo group [14] also detected this.
An X-ray transient XRF020427 caused sudden ionospheric disturbance (SID) in
NWC (19.8 kHz) signal at Perth, Australia [15]. 
Another SID caused by the prompt X-rays and/or $\gamma$ rays from  the
GRB 030329 ionizing the upper atmosphere was detected at Kiel, Germany [16]. 
On December 27, 2004, at $\sim$21:30:26 UT a giant hard X-ray/$\gamma$ ray flare from the 
soft gamma ray repeater SGR 1806-20 was also detected [17]. 

Given that only a handful of detection have been reported so far, we believe that our
positive detection of at least two more objects is significant. In the next Section,
we present our results briefly. Finally in Section 3, we give our concluding remarks.

\section {ICSP Monitoring of an SGR and a GRB}

Indian Centre for Space Physics has several receiving stations of Very Low Frequency (VLF)
radio signals which primarily observe VTX station (18.2KHz) of Indian Navy located near the southern end of 
Indian sub-continent ($08^\circ23'$N, $77^\circ45'$E). The data is automatically stored in the computer through the 
data acquisition card and software. Figure 1 shows the locations of the VTX transmitter and the receivers  
at Malda, Salt Lake, ICSP(Garia) and Pune. Notably, the VTX station is located near the magnetic equator
and almost vertical magnetic meridian (150$^\circ$E) divides the Indian subcontinent into two halves.

\subsection{SGR J1550-5418}

At 00:53:52 UT on 22nd January, 2009 SGR J1550-5418 started flaring and was detected by FERMI 
and subsequently by several satellites (e.g., [18]). The source bursted many
times on that day and were simultaneously detected in VLF signals [7-8]
for the first time for this object.  Figure 2 shows the location of the subflare point
when it bursted first and illuminated the Indian sky. There were
358 satellite detections reported within the time the source was over Kolkata sky. However,
many of them were very weak or too close to resolve. We clearly resolved 
$73$ detections which coincided with the satellite detection timings (within a few seconds
compatible with the ionization time scale). Figure 3 shows the VLF signal amplitude (as observed)
as a function of time over a few seconds (in UT on 22nd January, 2009). The long dashed lines are 
reported satellite observations while the dotted dashed lines are the corresponding perturbations in the amplitude. 
Very often successive events occurred even before the earlier one decayed. 
It can be easily shown that each fast rise and exponential decay (FRED) signal can be fitted 
well by Kocevski et al. [19].
$$
F(t)=F_{m}(\frac{t}{t_{m}})^{r}[\frac{d}{d+r}+\frac{r}{d+r}(\frac{t}{t_{m}})^{(r+1)}]^{-(r+d)/(r+1)},
$$
where, $F_{m}$ is the maximum flux at $t_{m}$, $r$ and $d$ are the rising and decaying indices, 
respectively. 

\subsection {GRB 090424}

The Gamma Ray Burst GRB 090424 occurred on 24th April, 2009 at 14:12:09 UT. Many satellite observations 
were reported (e.g., [20, 21]). According to FERMI in 8-1000 keV band, the fluence is $5.2 \times 10^{-5}$ ergs/cm$^2$.
The sub-flare point was located in the north of Philipines. 
This event produced a measurable VLF disturbance in the signal received at three stations located at
ICSP (Garia), Salt Lake ($22^\circ$34'N , $88^\circ$24'E) and Malda ($25^\circ$00'N , $88^\circ$09'E) 
all located in the eastern side of India [9]. 
ICSP and Salt Lake stations are only 10 miles away. The Malda station is $\sim$200 miles 
away from the former two stations. VTX signal in Pune ($18^\circ$34'N , $73^\circ$49'E)
did not respond to this GRB. This station is located in the west of Kolkata by $1000$ miles. 

The vertical line in Fig. 4  is the satellite detection time of the GRB (in UT on 24th April, 2009). 
The upper panel is the RHESSI satellite observation (background subtracted).
The VLF response occurred after a few seconds. The signals in all the three places were found to be 
correlated immediately after the event for tens of seconds.

\section{Conclusions}

The Earth's ionosphere is a large, open and noisy detector, but it is free of production and maintenance costs.
We reported here two observations where we showed how the ionosphere responded to 
one Soft Gamma Ray repeater and one Gamma Ray Burst. Given that only a handful of such detections are 
present in the literature, our observation certainly is a significant addition. From a careful
analysis, significant knowledge about the interaction of high energy radiation with 
the Earth's ionosphere can be obtained. The analysis is in progress and will be reported elsewhere.

From the astrobiological point of view, the burning of the ionosphere by the constant bombardments
of the extra-terrestrial energetic events is of great concern. It has been reported by Inan et al. [22]
that the otherwise neutral atmosphere at a height of $30$ km was ionized by the SGR 1806-20. If 
any such event happened in our own galaxy, the effect could have been devastating and the
whole atmosphere would have been converted to ionosphere for a short-while at least, exposing
us to hostile radiations from space. In this sense, our study is linked to the exobiology 
and more such studies are encouraged in this direction.

\section*{Acknowledgments}

SKM acknowledges the support from a CSIR/JRF Fellowship. SS acknowledge the support from ISRO
RESPOND project.

 {}

\begin{figure}
\vbox{
\centerline{\psfig{figure=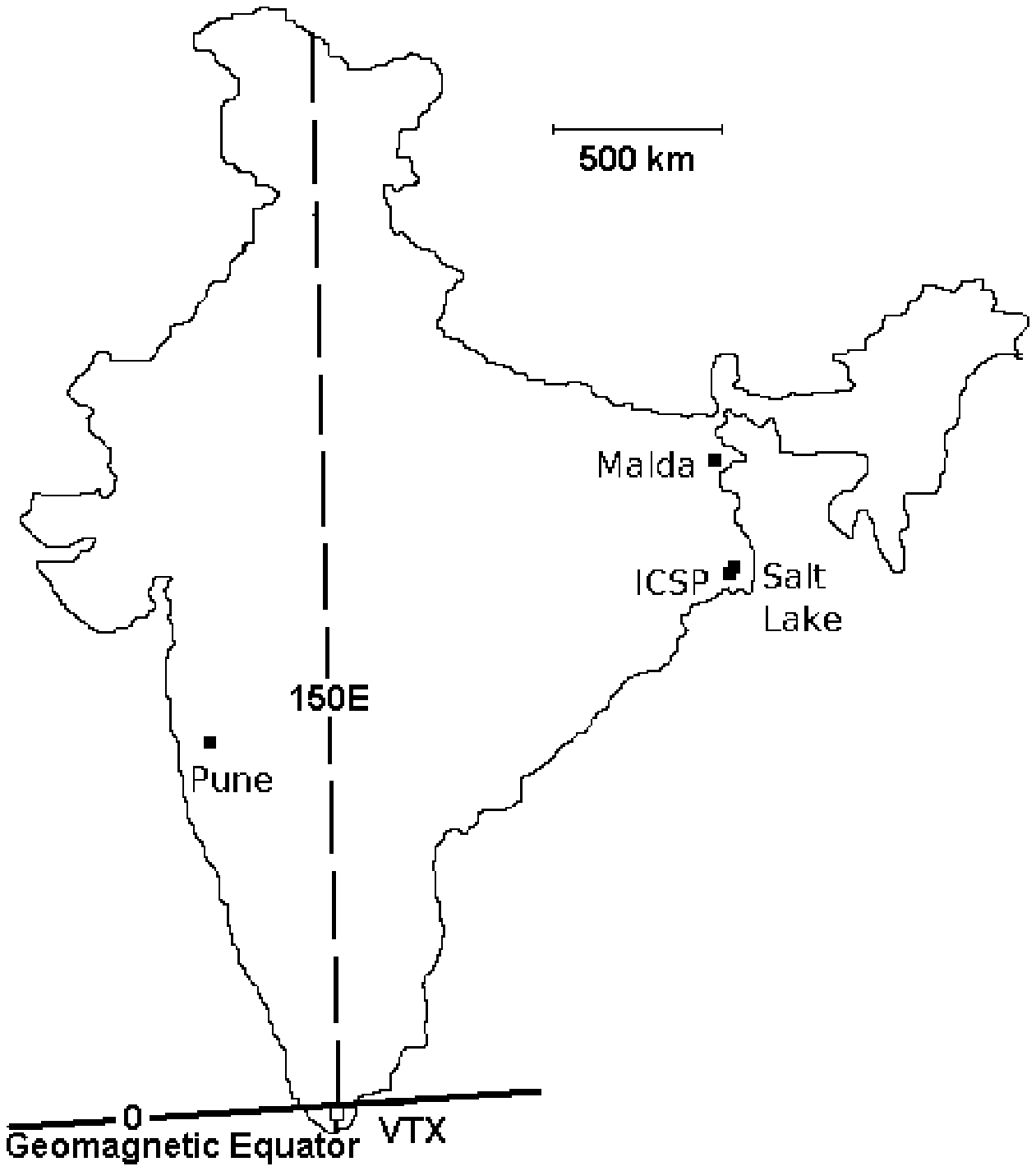,height=2.0in}}}
\noindent{{\bf Figure 1.} The locations of the transmitter (VTX) and receivers (Malda, Salt Lake, ICSP, Pune). }
\end{figure}

\begin{figure}
\vbox{
\centerline{\psfig{figure=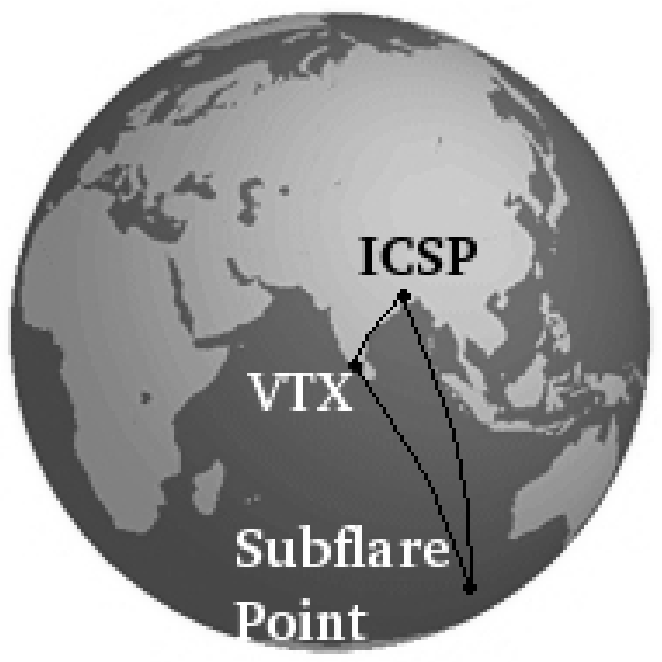,height=3.0in}}}
\noindent{{\bf Figure 2.} The sub-flare point of the SGR at the time of onset 
is shown {\it vis-\'a-vis} the VTX transmitter at southern tip of India and Kolkata receiving station.}
\end{figure}

\begin{figure}
\vbox{
\centerline{\psfig{figure=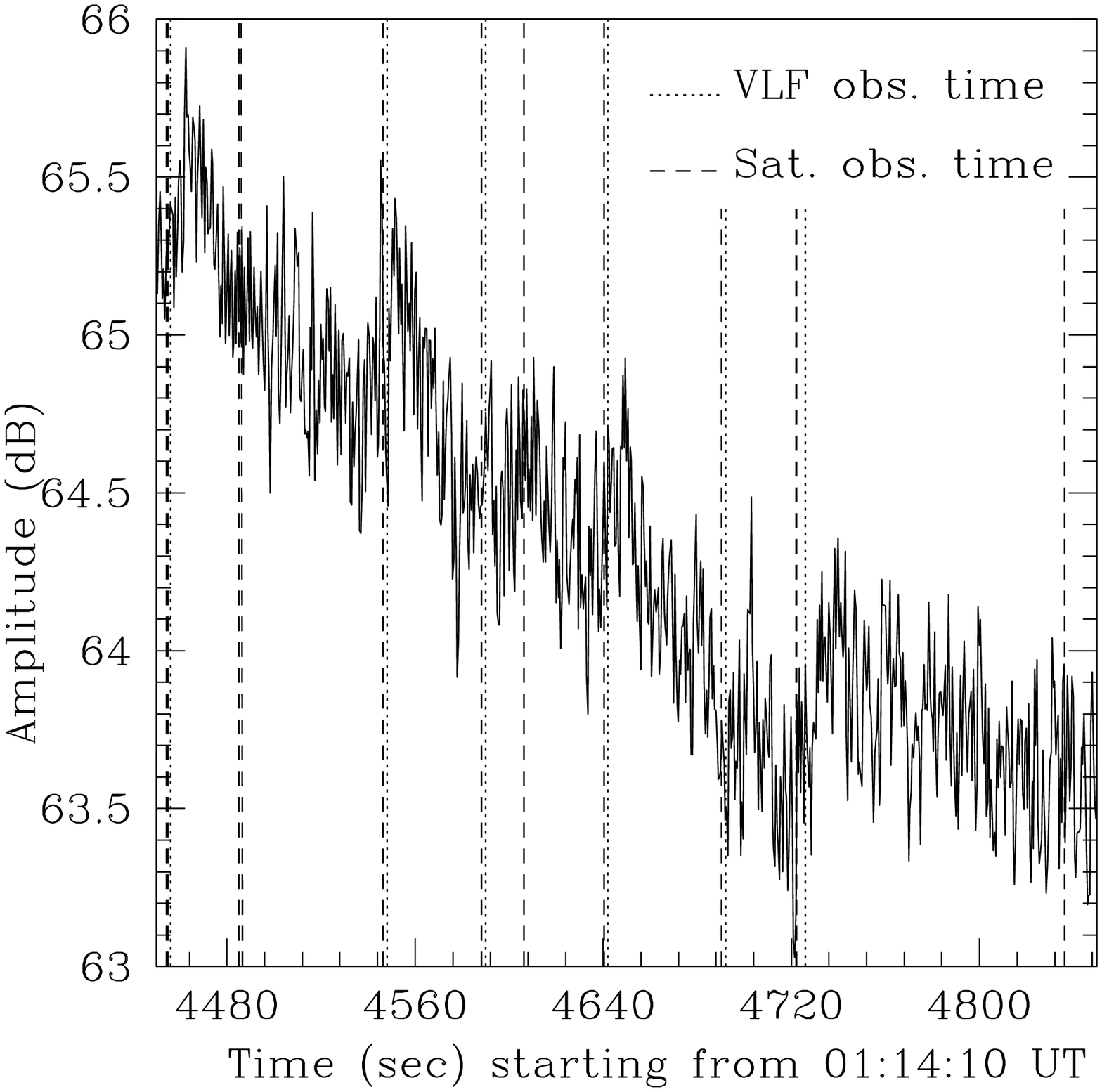,height=3.0in}}} 
\noindent{{\bf Figure 3.} VLF amplitude obtained at Salt Lake (UT on 22/1/2009) station. 
Satellite events are shown by vertical dashed line and the VLF events are shown by dotted line.}
\end{figure}

\begin{figure}
\vbox{ 
\centerline{\psfig{figure=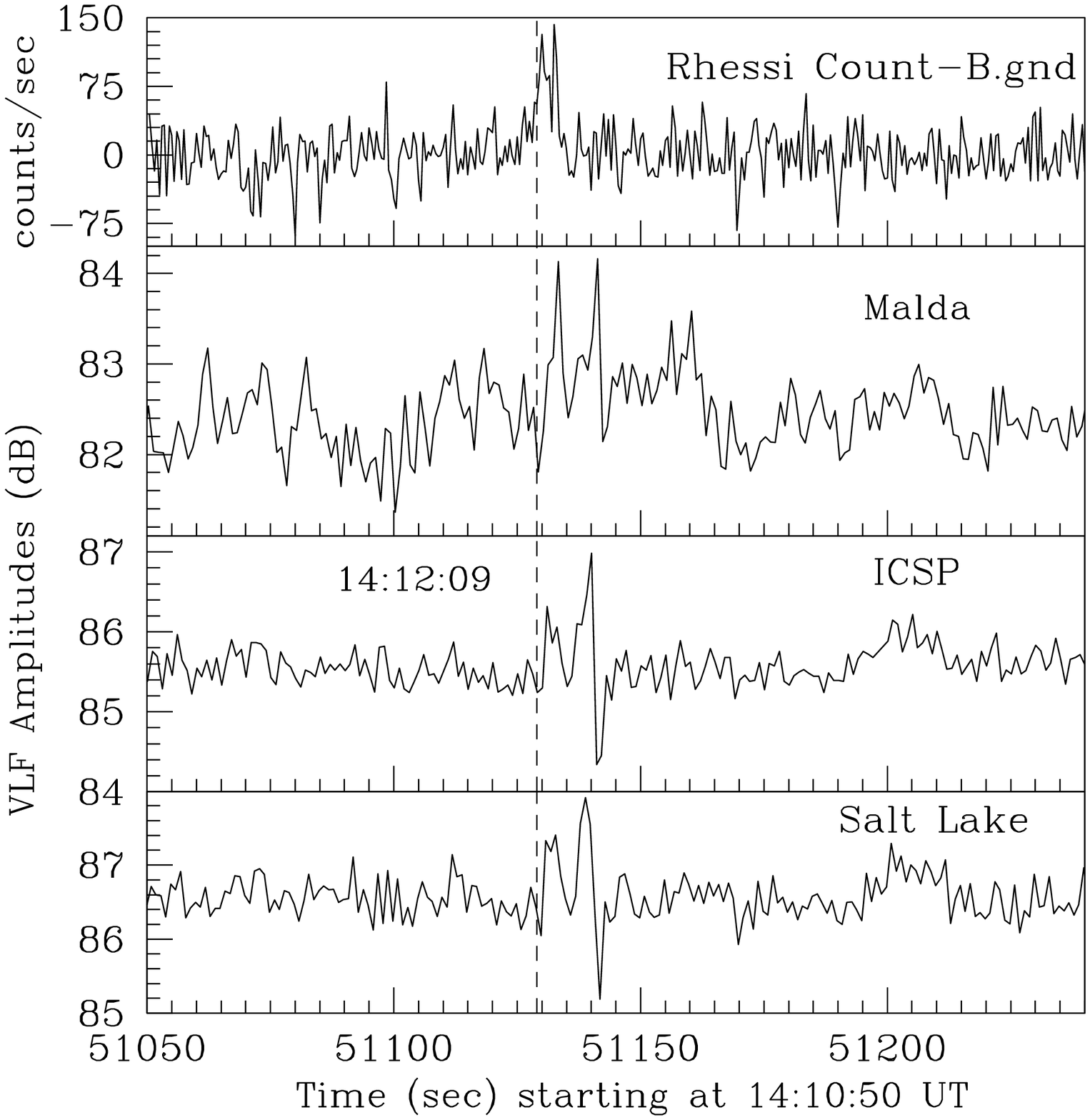,height=3.0in}} }
\noindent{{\bf Figure 4.} At example of the satellite observation of GRB090424 by RHESSI satellite (upper panel) and 
the perturbations of the VLF signals at different stations (lower panels).}
\end{figure}

\end{document}